\documentclass{birkjour}
\usepackage{gensymb}
\usepackage[usenames,dvipsnames]{color}
\usepackage{tikz}
\usepackage{epsfig}
 \newtheorem{theorem}{Theorem}[section]
 
 \newtheorem{lemma}[theorem]{Lemma}
 
 \theoremstyle{definition}
 \newtheorem{definition}[theorem]{Definition}

 \numberwithin{equation}{section}

\begin{document}

%
%
%
%
%
%
%
%
%

\title[Moffatt vortices: Concerns and Finiteness]
 {Moffatt vortices: Concerns and Finiteness}

\author[Jiten C. Kalita]{Jiten C. Kalita}

\address{%
Department of Mathematics\\
Indian Institute of Technology Guwahati\\
PIN 781039, INDIA
}

\email{jiten@iitg.ernet.in}

\author{Sougata Biswas}
\address{Department of Mathematics\\
Indian Institute of Technology Guwahati\\
PIN 781039, INDIA}
\email{b.sougata@iitg.ernet.in}
\author{Swapnendu Panda}
\address{Department of Mathematics\\
Indian Institute of Technology Guwahati\\
PIN 781039, INDIA}
\email{p.swapnendu@iitg.ernet.in}



\begin{abstract}
Till date, the sequence of vortices present in the solid corners of steady internal viscous incompressible flows, widely known as Moffatt vortices was thought to be infinite. However, the already existing and most recent geometric theories on incompressible viscous flows that express vortical structures in terms of critical points in bounded domains, indicate a strong opposition to this notion of infiniteness. In this study, we endeavor to bridge the gap between the two opposing stream of thoughts by addressing what might have gone wrong and pinpoint the shortcomings on the assumptions of the existing theorems on Moffatt vortices. We provide our own set of proofs for establishing the finiteness of the sequence of Moffatt vortices by making use of the continuum hypothesis and Kolmogorov scale, which guarantee a non zero scale for the smallest vortex structure possible in incompressible viscous flows. We point out that the notion of  infiniteness resulting from discrete self-similarity of the vortex structures is not physically feasible. The centers of these vortices have been quantified by us as fixed points through Brouwer fixed-point theorem and boundary of a vortex as circle cell. With the aid of these new developments and making use of some existing theorems in topology along with some elementary concept of mathematical analysis, we provide several approaches to delve into this issue. All these approaches converge to the same conclusion that the sequence of Moffatt vortices cannot be infinite; in fact it is at most finite.
\end{abstract}

\maketitle
\section{Introduction}
Terms like incompressible, inviscous, inviscid, impermeable etc. have assimilated into the vocabulary of fluid mechanics quite naturally over the years \cite{batch, white}. The usage of these words seem to be very casual in the context of  the desired precision or accuracy in the field of Mathematics and Physics, yet one has hardly seen any resistance from the scientific community over the usage of such words. This is probably because of the fact that all these words have certain quantification. For example, although no flow in the real world is incompressible, this word has wide acceptability amongst the fluid dynamics community for fluid flows that remain within a Mac number limit of 0.3 \cite{batch, neu}. However, it is highly questionable whether  one can apply the word {\it infinite} in the same vein to mean a large number of vortices present in a fluid flow without actually quantifying it, which of course is an extremely difficult task.

In the study of fluid flow, vortices are very important structures as they play a big role in controlling the dynamics of a flow \cite{banks}. Though they are known to occur in the vicinity of solid walls where the flow separation takes place, scientists have always been keener on probing the nature of vortices at the corners of solid structures. The existence of a sequence of vortices at the corner of solid structure for internal flows with decreasing size and rapidly decreasing intensity has been indicated by several physical experiments and mathematical asymptotics \cite{moff1, moff2, moff4, moff3, taneda}.  Their formation, evolution and progression on different geometrical and physical set up have always generated a lot of interest  both amongst   fluid dynamicists as well as mathematicians.

Study on the existence of these vortices dates back to the pioneering work of Dean \& Montagnon \cite{dean} which was later consolidated by Moffatt \cite{moff1, moff2}. The  flow near a sharp corner between two bounding planes was considered and a solution of the purely biharmonic equation for  creeping flows \cite{tasos} was sought  in terms of some exponential power $\Lambda$ of the distance $\tilde{r}$ from the corner. He found this $\Lambda$ to be a complex number when the angle between the two planes is less than about $146\degree$, implying infinite oscillations and hence concluded that there exists an infinite sequence of counter-rotating eddies as one approaches the corner. From then onwards, the occurrence of such corner vortices, named aptly as ``{\it Moffatt vortices}" has always been synonymous with the existence of an infinite sequence, albeit without any rigorous mathematical proof.

In 1976, Collins \& Dennis \cite{col} computed the flow of a slow viscous fluid in a curved tube having a cross-section in the shape of a right-angled isosceles triangle. They observed
vortices of Moffatt's type firstly in the secondary flow in the $45\degree$ corners and then in the 90$\degree$ corner. By refining the grid size of the computational domain and making use of extrapolation technique, they were able to trace thirteen  vortices in the $45\degree$ corner and six pairs of vortices in the 90$\degree$ corner. In 1979, Taneda \cite{taneda} tried to establish the existence of these vortices experimentally in a V-notch. However, the visualization of such vortices revealed the existence of only a few of them in the corner. This was followed by several theoretical and numerical experiments \cite{gustafson3,hellou,liron,robert,moff4,moff3,oneill,poz} on vortices around a corner in solid structures.

Of late, there has been a surge of theoretical and numerical studies on corner vortices  \cite{biswas,deshpande,gustafson,hall,kirk,kras,kras1,malyuga,  shankar1,shankara,shtern} for slow viscous flows on different geometries. In 1993, Anderson \& Davis \cite{anderson}
presented a local picture of steady, two-dimensional (2D) viscous flow of two fluids in a wedge. They figured out the geometries for which wedge angle solutions exist, and also identified conditions under which Moffatt vortices may appear in the flow. During 1998-2000, Shankar and Deshpande \cite{shankara}, and Deshpande and Milton \cite{deshpande} studied their existence in the 2D lid driven cavity flow. In 2004, Biswas {\em et al.} \cite{biswas} investigated laminar backward-facing step flow for a wide range of Reynolds number and based on the theory of Moffatt \cite{moff1,moff2}, concluded that an infinite sequence of closed eddies with decreasing size and strength is expected for $Re \to 0$.  It is worth mentioning that although the study of Moffatt vortices mainly pertains to Stokes flow, the  sequence of corner vortices in decreasing size and intensity can be found for moderately high Reynolds number flows as well. This is well documented in the enormous amount of numerical and experimental studies being carried on the lid-driven cavity flows \cite{albe, jiten16, deshpande, koseff, sah, shankara, shankara2, sousa}.

In 2005, Malhotra \emph{et al.} \cite{chetan} reviewed the Moffatt problem \cite{moff1} and established the existence of Moffatt vortices for the Stokes flow bounded by two concentric coaxial cones with a common vertex through eigenvalue analysis and described the asymptotic distribution of eigenvalues for both even and odd flow structures. In the same year, Malyuga \cite{malyuga} considered Stokes flow in a circular cone driven by a non-zero velocity applied to the boundary within the ring which are represented in the form of Fourier series. He obtained the transcendental equation for the eigenvalues, which determines the asymptotic behavior of the flow in the neighborhood of the vertex, arriving at a conclusion similar to that drawn by Moffatt \cite{moff1,moff2}.
In 2014, Kirkinis \& Davis \cite{kirk}, by using a hydrodynamic theory of liquid slippage on a solid substrate concluded that an infinite sequence of vortices is formed in a moving liquid wedge of certain angle between a gas-liquid interface and a rigid boundary. Other recent studies in this direction worth mentioning are \cite{hall,kras,kras1,shankar1,shtern}.

All the existing geometrical theories on incompressible viscous flows (Bakker \cite{bakker}, D\'{e}lery \cite{delery}, Ma \& Wang \cite{wang}, Hirschel \cite{hir}, Wu {\em et al.} \cite{wu}) based on the concept of dynamical system \cite{perko} express vortical structures in terms of critical points in bounded domains. They again indicate a strong opposition to the notion of infiniteness. The presence of a sequence of vortices  at the corner of decreasing size and intensity has already been emphatically established by laboratory and numerical experiments \cite{biswas, jiten16, col, deshpande,gustafson,taneda}. However, neither any authentic mathematical proof nor any laboratory experiment till date has been able to establish their infiniteness. Despite this, mathematicians and engineers alike went on producing a large number of works on the topic of corner vortices
propounding their infiniteness, as evidenced by many of the works cited above. Even for the lid-driven cavity problem, which over the years has become the most frequently used benchmark problem amongst the computational fluid dynamics community and is an obvious example of internal incompressible viscous flow, claims of infinite sequence of corner vortices can still be found in the existing literature for this problem \cite{albe, deshpande, koseff, sah, shankara, shankara2, sousa}.

No one paid much attention to this issue untill Gustafson {\it et al.} \cite{gustafson2} in 1989 tenderly questioned on the issue of the infiniteness of Moffatt vortices. They concluded that the computational resources available at that time was not enough to provide a conclusive answer to this question (more details in section \ref{concern}). As such the task of establishing the finiteness of the sequence of such vortices still remained unaccomplished and the question on infiniteness unanswered. The objective of the current study is to explore a possible missing link between the contrasting theories of Moffatt vortices \cite{anderson, biswas, branicki,col, deshpande,hall,hellou,kirk,kras,kras1,liron,chetan,malyuga,
robert, moff1,moff2,moff4,moff3,oneill,poz,shankar1,shtern} and the recent geometrical theories on incompressible viscous flows \cite{bakker,delery,ghil, hir,jfmkalita,wang,oba,wu}. We endeavor to bridge the gap between the factions by pinpointing what could have possibly gone wrong with the assumptions of the existing theories upon which the conclusion of infiniteness is built.

Right from the time the term Moffatt vortices \cite{moff1,moff2} has been coined, their study on various geometries has mostly been through the eigenvalue analysis. To the best of our knowledge the study of Moffatt vortices under the purview of their topology and critical point theory has never been carried out before. In the current study, we for the first time, establish some novel theories on Moffatt vortices including
\begin{itemize}
\item[(i)] quantification of centers of these vortices as fixed points (see section \ref{center}, lemma \ref{bft_vort}) through Brouwer Fixed-Point Theorem (Theorem \ref{bft}),
\item[(ii)] quantification of the largest neighborhood of the fixed point of a particular vortex as circle cell by extending the idea of circle cell for divergence-free vector fields of Ma \& Wang \cite{wang}.
\end{itemize}

Based on the above theories and some recent developments in geometric theory of incompressible viscous flows \cite{delery,wang}, and making use of some elementary mathematical analysis \cite{rudin}, we  prove that the sequence of Moffatt vortices in fluid flows around solid corners is  finite. It is worth mentioning that the same conclusion is reached by tackling the hypothesis through six different approaches. Note that in all the studies mentioned above, two-dimensional flows as an idealization of a three-dimensional (3D) one or 3D flows having symmetry in one direction were considered.  As such all the theories developed by us will be considered over planner regions only.

 The paper is arranged in the following manner: Section \ref{contro} deals with the controversies with the issue of the infiniteness of Moffatt vortices,  Section \ref{wrong} pertains to a brief on what might have gone wrong with the theory,  Section \ref{sec3} with  preliminaries for proving the finiteness of Moffatt vortices,  Section \ref{sec5} with the proofs on the finiteness of Moffatt vortices and finally Section \ref{sec6} summarizes the whole work.
\section{The controversies}\label{contro}
In this section, we throw some lights on the controversies surrounding the notion of infiniteness. They include the mathematical origin of Moffatt vortices, certain concerns and questions over the issue which have not been settled till date.

\subsection{The mathematical origin of the infiniteness of Moffatt vortices}\label{origin}
The existence of vortices in a flow field is a highly nonlinear phenomena and
most importantly there exist an inherent connection between their occurrence
and the nonlinear nature of Navier-Stokes (N-S) equations \cite{jiten16}.
However, the theoretical studies on Moffatt vortices seek the solution of the biharmonic form of the steady-state N-S equations for Stokes flow in the streamfunction $\psi$ as
\begin{equation}\label{bhr}
\nabla^4\psi=0,
 \end{equation}
which is a linear one. Thus it completely annihilates all the effects of nonlinearity. 

The solution of (\ref{bhr}) in polar coordinates $(\tilde{r},\phi)$ is assumed to be of the form $\displaystyle \psi=\tilde{r}^\Lambda f(\phi)$ leading to an equation in $f$
 \begin{equation}
 f''''+\left\{(\Lambda+1)^2+(\Lambda-1)^2\right\}f''+
 (\Lambda+1)^2(\Lambda-1)^2f=0,
 \end{equation}
 resulting in a solution of the form \cite{moff1}
 \begin{equation}
 \label{sln}
 f(\phi)=C_1\sin (\Lambda-1)\phi+C_2\cos (\Lambda-1)\phi+C_3\sin (\Lambda+1)\phi+C_4\cos (\Lambda+1)\phi .
 \end{equation}
 The existence of the infinite sequence of vortices follows from ones ability in arriving at a solution of the form (\ref{sln}). Making use of (\ref{sln}), one can find the transverse component of velocity on the plane $\phi=0$ as \cite{moff1}
  \begin{equation}\label{vel_tran}
 v_{\phi=0} \sim \gamma \frac{1}{\tilde r}\bigg(\frac{\tilde r}{\tilde r_0}\bigg)^{p_1+1}\sin\bigg(q_1\mbox{ln}\frac{\tilde r}{\tilde r_0}+\tilde \epsilon\bigg).
 \end{equation}
 The notion of an infinite sequence of vortices near the corner comes from the above expression by concluding that it changes sign infinitely as $\tilde {r} \to 0$.
 
Another argument for the existence of the infinite sequence of eddies near a solid corner comes from the concept of discrete self-similarity. Consider a flow domain $\Omega=\{(x,y)\in \mathbb{R}^2 | x=\tilde r\cos \phi,\; y=\tilde r\sin \phi,\; \tilde r>0, \; |\phi|<\alpha\}$, in which these vortices are defined in the flow whose streamfunction $\psi$ has the form
$$\psi(x, y)=\mbox{Re}\bigg(\tilde{r}^{\Lambda}\{A\cos \Lambda \phi + B\cos(\Lambda-2)\phi\}\bigg),$$ where $A,\; B$ are some nonzero constants which satisfy $$\sin2(\Lambda-1)\alpha + (\Lambda-1)\sin 2\alpha=0.$$
Then, if one considers the scaling $$\psi_{\Lambda_n}(x,y)=\Lambda_n^{-(\mathrm{Re}(\Lambda-1)+1)}\psi(\Lambda_n x,\Lambda_n y),$$
both the domain $\Omega$ and streamfunction $\psi$ remain invariant under the action of the scaling $(x,y) \to (\Lambda_n x,\Lambda_n y)$, where $\displaystyle \Lambda_n=\frac{2 \pi n}{|\mbox{Im}(\Lambda -1)|},~n \in \mathbb{N};~\mbox{i.e,}~\psi_{\Lambda_n}(x,y)=\psi(x, y)$. Note that the concept of discrete self-similarity follows from this notion of invariance. 


\subsection{The concerns over infiniteness}\label{concern}
Probably the earliest concern over the infiniteness of Moffatt vortices was raised by Gustafson {\em et al.} in  the year 1989 \cite{gustafson2} where they questioned: ``{\it Are any of these entities truly infinite dimensional?}";  by {\it these entities}, they meant the corner vortices in the famous lid driven cavity flow. According to them, what actually had been sought physically or mathematically, was their existence in a large finite dimensional dynamical system without further venturing into the metaphysical meaning of infinity. They further mentioned that though the theory predicts an infinite sequence of vortices at the corner of the solid structure (in particular they considered the lower two corners of the driven cavity), it is a linearized one (see equation (\ref{bhr})). They had reservations on how much it depends on the linearizing assumptions and for what range of Reynolds numbers it continues to hold for the full nonlinear N-S equations, which are the governing equations for  incompressible viscous flows \cite{gustafson2}. True to their concerns,
many of the existing and recent theorems on separation of incompressible viscous flows \cite{bakker, delery, jfmkalita, wang, hir, wu}  lean towards the existence of a finite number of vortices in a finite domain including corners.

In their study of unsteady separation induced by a vortex, Obabko and Cassel \cite{oba} discusses about the viscous-inviscid interaction leading to spike formation. The presence of a primary vortex induces an adverse pressure gradient along a solid surface and the aforesaid interaction accelerates the spike formation leading to the formation of secondary vortices. The mechanisms for the creation of the tertiary, quaternary and the succeeding vortices is the same \cite{jfmkalita}.   The {\it structural bifurcation theory} of Ghil {\it et al.} \cite{ghil}, by predicting the exact location and time of the birth of a vortex  clearly establishes that the birth of the vortices in the sequence in a corner occurs one after another in succession. The clear implication of all these theories \cite{wang} is the following fact: the birth of two vortices in succession or any two vortices in the same sequence cannot take place at the same instant of time.
 This is in direct contrast with the infiniteness of corner vortices in steady state flow resulting from the solution of (\ref{bhr}), which would have taken infinite time to reach the steady state through time-marching.

All the existing geometrical theories on incompressible viscous flows (Bakker \cite{bakker}, D\'{e}lery \cite{delery}, Ma \& Wang \cite{wang}, Hirschel \cite{hir}, Wu {\em et al.} \cite{wu}) based on the concept of dynamical system \cite{perko} express vortical structures in terms of critical points in bounded domains. They again indicate a strong opposition to the notion of infiniteness. Despite this, even for the lid-driven cavity problem, which over the years has become the most frequently used benchmark problem amongst the computational fluid dynamics community and is an obvious example of internal incompressible viscous flow, claims of infinite sequence of corner vortices can still be found in the existing literature for this problem \cite{albe, deshpande, koseff, sah, shankara, shankara2, sousa}.

It is a well known fact that the formation of the so called infinite sequence of vortices in Stokes flow is due to 
the effect of certain stirring/rotating force far from the corner \cite{anderson, branicki, col, hall, hellou, kirk, kras, liron, chetan, malyuga, robert, moff1, moff2, moff4, moff3, oneill, shankar1, shtern}. A further undermining into the existing literature reveals a very vague picture of the extent of the domains over which the flow is considered. If the source of the force is an infinite distance away from the corner, the effect of stirring/rotating force will fall down gradually with a proportional fall in the intensity of the force or the strength of the vortices at their centers as one moves away from the source. At a certain distance from the source of the given force, no effect of it will be felt. Thus the process of the formation of the vortices  will be well over much before reaching the corner. On the other hand the theory of Moffatt vortices considers the existence of the infinite sequence vortices in the neighborhood of the corner where $\tilde{r} \to 0$. Thus a source force at an infinite distance from the corner  nullifies the presence of the so called infinite sequence of vortices at the corner indicated by mathematical asymptotes.

One of the main source of the infiniteness of the sequence of vortices is the so called discrete self-similarity of infinite degree of these vortices, which unfortunately is not physically feasible. For example, natural objects like fern and cauliflower exhibit self-similarity only under finite degrees of magnification (upto finite number of stages/steps). Moreover, if this notion of infinite degree of magnification is to hold true, even in Stokes' flow,  one must be able to actually accommodate an entire ``tail" of eddies-sequence below the Kolmogorov (see section \ref{kol}) length-scale, which is physically impossible.
\subsection{The unanswered questions}\label{big}
Only recently, in 2006, Moffatt and his co-worker Branicki \cite{branicki} has broached upon the possibility of the finiteness of these sequence of vortices for certain cases. Analyzing the time-periodic evolution of Stokes flow near a corner, they concluded that depending upon the smoothness and angle of the corner, and on  the nature of the forcing, an infinite sequence of corner eddies may be present if the corner is sharp. On the other hand, if corners are rounded off so that the boundary is everywhere analytic, it is expected that a finite sequence of eddies may still form in regions near points of maximum curvature on the boundary. But the big question here is:
Is there a slight transition from a smooth  to a sharp corner  good enough to trigger infiniteness to the sequence of vortices? If so, how does one quantify this sudden jump from finite to infinite number of vortices and then again, what is this thin line between the extent of smoothness and sharpness leading to this enormous jump?

Moreover, as mentioned in section \ref{origin}, the concept of Moffatt vortices comes from the solution of the linearized version of the Navier-Stokes equations. These equations are derived under the assumption of ``{\it Continuum Hypothesis}" \cite{batch,white}. According to this hypothesis, even the smallest volume scale cannot be zero; for example, for air, it is of the order $10^{-9}{\rm mm}^3$ containing approximately $3 \times 10^7$ molecules under standard conditions.

Note that an infinite sequence of vortices of decreasing size renders a size zero to the vortices belonging to the tail of the sequence (see Proof 4 of Theorem 7.1 in section \ref{sec5}). However, this conclusion stems out from the solution of equation (\ref{bhr}) which is built under the assumption of continuum hypothesis requiring a minimum non-zero volume scale. This clearly contradicts the existence of an infinite number of vortices in the corner of solid structures. The Kolmogorov length scale corroborates this fact.

\subsubsection{Kolmogorov (length scale) theory}\label{kol}
Stokes flow and turbulent flows are at the extreme ends of the spectrum of incompressible viscous flow regime characterised by the Reynolds number. Therefore, the mention of the Kolmogorov scale \cite{les,loh,pkm,pkm2,pope} may sound totally irrelevant in the context of Stokes flow. Juxtaposed to this, this length scale plays an important role in our analysis. The Kolmogorov theory clearly states that vortices cannot exist below a certain non-zero length scale \cite{deshpande,wal},  since the local value of power density ($\varepsilon$) would be so high that the kinetic energy would be fully dissipated as heat.


An estimate for the scales at which the energy is dissipated is based only on the dissipation rate and viscosity. If the dissipation rate per unit mass ($\varepsilon$) has  dimensions ($m^2/sec^3$) and viscosity, $\nu$ has dimension ($m^2/sec$) then the length scale formed from these quantities is given by
$$\eta=\bigg(\frac{\nu^3}{\varepsilon}\bigg)^{1/4}.$$
This length scale is called the Kolmogorov length scale \cite{andersson, fris, jim}.

Note that the smallest length scale for incompressible viscous flows occurs in the turbulent regime. Stokes flow, for that matter laminar flows in the moderate Reynolds number regimes will have vortices having much bigger scales than those prevalent in turbulent regime. As such, the size of a vortex under consideration in the current study cannot fall below  the Kolmogorov length scale.

\section{The notion of infiniteness: What might have gone wrong?}\label{wrong}
In the above, we listed the concerns and the related questions in connection with the infiniteness of Moffatt vortices. In the following, we endeavor to bridge the gap by addressing what actually went wrong with the notion of infiniteness and subsequently providing our own set of proofs on the finiteness of corner vortices.
Here we will reflect upon the shortcomings of the assumptions of the existing theorems upon which the conclusion of infiniteness of corner vortices is built. These observations will help us pinpoint where the existing hypothesis went wrong and pave the way for providing a concrete mathematical basis that predict the correct physical phenomenon.

The foremost argument provided in favour of the infinite sequence of vortices stems out from the velocity component (\ref{vel_tran}) arising out of equation (\ref{bhr}).
 As discussed in sections \ref{origin}  and \ref{big}, (\ref{bhr}) inherently assumes the  ``Continuum Hypothesis", can this $\tilde r$ in the expression $$\sin \bigg(\log \frac{\tilde r}{\tilde r_0}\bigg) $$
actually tend to zero? Note that the absolute size of the eddies is directly proportional to the length scale $\tilde r_0$. For a large $\tilde r_0$, one may mathematically let $\displaystyle \frac{\tilde r}{\tilde r_0} \to 0$ by letting $\tilde r_0 \to \infty$, however, in reality $\tilde r$ can never tend to zero. Moreover on a finite domain, $\tilde r_0$ is bounded and hence can never tend to $\infty$. As such the assumption in \cite{moff1} that $\tilde r_0$ is an arbitrary length scale is physically incorrect. This is where the mathematical exuberance tends to overshadow  physical reality in a conflicting manner; ideally one would desire mathematics to go hand in hand with physics.

Furthermore, the number of eddies will  decrease even more due to dampening-effect after the collision of the fluid with the wall. This effect is completely ignored in this context. Note that the notion of infiniteness comes from the so called infinite oscillations of the sine waves depicting the streamfunction value as a solution of (\ref{bhr}). However, on a finite domain, the number of such waves must be finite and they can be thought as representing discrete oscillations \cite{breth}. The maximum wave length that can be accommodated inside a finite domain is the longest inscribable rectilinear length in the domain. The measure of the minimum wavelength is equal to twice the mean free path of the fluid molecules, which is nothing but the average distance traveled by a moving fluid particle between successive collisions. Thus even when the vortices reach molecular level, this minimum wavelength cannot be zero. As such the largest number of vortices having a one to one correspondence with the oscillatory waves must be finite.

\section{Preliminaries}\label{sec3}
In the following, we provide some definitions, theorems and results from the existing literature on topological fluid dynamics \cite{bakker, pollack, perko, ricca} and some of our own newly developed theories (see section \ref{center}), which will be used in appropriate junctures to prove the finiteness of the corner vortices.
\subsection{Notations}
\begin{itemize}
\item $\mathbb{R}$ is the set of real numbers.
\item $\mathbb{N}=\{1, 2, 3, \cdots\}$ is the set of natural numbers.
\item $\tilde{V} = \{q \colon q\in V_i$, the $i$-th vortex in the sequence of Moffatt vortices\}.
\item $M$ is a planar region.
\item On the boundary $\partial M$ of the region $M$, $\hat{\tau}$ denotes the tangential vector and $\hat{n}$ the normal vector.
\item $T_pM = \{w\;|\: w \;\mbox{is tangent to} \; M\;\mbox{at}\; p\}$.
\item $TM = \{(p, T_pM)\;|\;p \in M\}$ is the tangent bundle of $M$.
\item Assume $r \geq 1$ is an integer. Let $\mathcal{C}^r(TM)$ be the space of all $r$-th differentiable vector fields $v$ on $M$.
\item $\mathcal{C}^r_{\hat{n}}(TM)= \{v \in \mathcal{C}^r(TM)\;|\;v\cdot \hat{n} = 0$ on $\partial M\}.$
\item $\mathcal{D}^r(TM)= \{v \in \mathcal{C}^r_{\hat{n}}(TM)\;|\;\nabla\cdot v = 0\}.$
\item $\mathcal{B}_0^r(TM)= \{v \in \mathcal{D}^r(TM)\;|\;v = 0$ on $\partial M\}.$
\end{itemize}
\subsection{Some essential topology}
\begin{definition}
\emph{A mapping $f$ of an open set $U \subset \mathbb{R}^n$ into $\mathbb{R}^m$ is called smooth if it has continuous partial derivatives of all orders.}
\end{definition}
\begin{theorem}\label{bft}
{\bf (Brouwer Fixed-Point Theorem \cite{pollack}):}
Any smooth map $\Psi$ of the closed unit ball $B^n \subset \mathbb{R}^n$ into itself must have a fixed point; that is, $\Psi(x)=x$ for some $x\in B^n$.
\end{theorem}
\subsection{Geometric theory of viscous incompressible flows}
\begin{definition}{\bf \cite{wang}}
\emph{A point $p \in M$ is called a singular point of $v \in \mathcal{C}^r_{\hat{n}}(TM)$ if $v(p)=0$.}
\end{definition}
\begin{definition}{\bf \cite{wang}}
\emph{A singular point $p$ of $v \in \mathcal{C}^r_{\hat{n}}(TM)$ is called non-degenerate if the Jacobian matrix of $v$ at $p$, $J_v(p)$ is invertible.}
\end{definition}
\begin{definition}{\bf \cite{wang}}
\emph{A vector field $v \in \mathcal{C}^r_{\hat{n}}(TM)$ is called regular if all singular points of $v$ are non-degenerate.}
\end{definition}
\begin{definition}{\bf \cite{wang}}
\emph{An orbit $\{\Phi(x,t)\}_{t \in \mathbb{R}}$ is called a closed (periodic) orbit if there is a time $T_0>0$ such that for any $t \in \mathbb{R}$, $\Phi(x,t)=\Phi(x,t+T_0)$.}
\end{definition}
\begin{lemma}{\bf\cite{wang}}
 \emph{Let $v \in \mathcal{D}^r(TM)(r \geq 1)$. Then each non-degenerate singular point of $v$ is either a center or a saddle point. A non-degenerate singularity on the boundary $\partial M$ must be a saddle point.}
\end{lemma}
\begin{theorem}\label{sct1}
{\bf (Structural Classification Theorem I \cite{wang}):} Let $v \in \mathcal{D}^r(TM)(r \geq 1)$ be regular. Then the topological structure of $v$ (flow field in the context of the current study) consists of a finite number of connected components, which are of the following types:
\begin{itemize}
\item[\emph{(1)}] circle cells, which are homeomorphic to open disks,
\item[\emph{(2)}] circle bands, which are homeomorphic to open annuli,
\item[\emph{(3)}] ergodic sets, and
\item[\emph{(4)}] saddle connections.
\end{itemize}
\end{theorem}
\begin{definition}{\bf \cite{wang}}
\emph{
Let $u \in \mathcal{B}^r_0(TM) (r \geq 2).$
\begin{enumerate}
\item[(1)] A point $p \in \partial M$ is called a $\partial$-regular point of $u$ if $\displaystyle \frac{\partial u_{\hat{\tau}}(p)}{\partial n} \neq 0$; otherwise, $p \in \partial M$ is called a $\partial$-singular point of $u$.
\item[(2)] A $\partial$-singular point $p \in \partial M$ of $u$ is called non-degenerate if
$$\displaystyle \mathrm{det}\begin{bmatrix}
\frac{\partial^2u_{\hat{\tau}}(p)}{\partial \hat{\tau}^2} & \frac{\partial^2u_{\hat{\tau}}(p)}{\partial {\hat{\tau}} \partial \hat{n}}\\
\frac{\partial^2u_{\hat{n}}(p)}{\partial \hat{\tau} \partial \hat{n}} & \frac{\partial^2u_{\hat{n}}(p)}{\partial \hat{n}^2}\\
\end{bmatrix} \neq 0
$$
A non-degenerate $\partial$-singular point of $u$ is also called a $\partial$-saddle point of $u$.
\end{enumerate}
}
\end{definition}
\begin{lemma}\label{lm_bd}
{\bf \cite{wang}} \emph{Each non-degenerate $\partial$-singular point of $u \in \mathcal{B}_0^r(TM)$ is isolated. Therefore, if all $\partial$-singular points of $u$ on $\partial M$ are non-degenerate, then the number of $\partial$-singular points of $u$ is finite.}
\end{lemma}
\begin{definition}{\bf \cite{wang}}
\emph{Let $\Omega \subset M$ be a closed domain. A point $p \in \Omega$ of $v$ (flow field) is called $\Omega$-boundary saddle (half-saddle) if there are only three orbits connecting $p$ in $\Omega$.}
\end{definition}
\subsection{Limit cycle}
\begin{definition}{\bf \cite{perko}}
\emph{A} {\it limit cycle} \emph{is an isolated closed trajectory.} By {\it isolated}, it is meant that \emph{neighboring trajectories are not closed; they spiral either toward or away from the limit cycle. If all the neighboring trajectories approach the limit cycle, we say the limit cycle is} {\it stable} \emph{or} {\it attracting}. \emph{Otherwise the limit cycle is} {\it unstable}.
\end{definition}
\begin{definition}{\bf \cite{perko}}
\emph{Consider the autonomous system}
\begin{equation}\label{sys1}
 \mathbf{\dot{x} = f(x)}
\end{equation}
 \emph{with $\mathbf{f(x)} \in \mathcal{C}^1(E)$ where $E$ is an open subset of $\mathbb{R}^2$.
 A point $\mathbf{\tilde p} \in E$ is an} $\omega$-{\it limit point} \emph{of the trajectory $\phi(\cdot,{\mathbf x})$ (which is a function from $\mathbb{R}$ to $E$) of the system (\ref{sys1}) if there is a sequence $t_n \rightarrow \infty$ such that $\displaystyle{\lim_{n \rightarrow \infty}{\phi(t_n,{\mathbf x})} = \mathbf{\tilde{p}}}$. Similarly, if there is a sequence $t_n \rightarrow -\infty$ such that $\displaystyle{\lim_{n \rightarrow \infty}{\phi(t_n,{\mathbf x})} = \mathbf{\tilde{q}}}$, then the point $\mathbf{\tilde{q}}$ is called an} $\alpha$-{\it limit point} \emph{of the trajectory} $\phi(\cdot,{\mathbf x})$.
\end{definition}
\begin{definition}{\bf \cite{perko}}
\emph{The set of all $\omega$-{\it limit points} of a trajectory $\Gamma$ is called the} $\omega$-{\it limit set} \emph{of $\Gamma$ and it is denoted
  by $\omega(\Gamma)$. The set of all} $\alpha$-{\it limit points} \emph{of a trajectory $\Gamma$ is called the} $\alpha$-{\it limit set} \emph{of $\Gamma$ and it is denoted by} $\alpha(\Gamma)$.
\end{definition}
\begin{definition}{\bf \cite{perko}}
\emph{A} {\it limit cycle} $\Gamma$ \emph{of a planar system is a closed solution curve (cycle) of (\ref{sys1}) which is the $\alpha$ or} $\omega$-{\it limit set} \emph{of some trajectories of (\ref{sys1}) other than $\Gamma$. If a cycle $\Gamma$ is the} $\omega$-{\it limit set} \emph{of every trajectory in some neighborhood of $\Gamma$, then $\Gamma$ is called an} $\omega$-{\it limit cycle} \emph{or} {\it stable limit cycle}; \emph{and if $\Gamma$ is the $\alpha$-{\it limit set} of every trajectory in some neighborhood of $\Gamma$, then $\Gamma$ is called an}  $\alpha$-{\it limit cycle} \emph{or} {\it unstable limit cycle}.
\end{definition}
\begin{theorem}\label{dulac}
{\bf(Dulac \cite{perko}):} In any bounded region of the plane, a planar analytic system (\ref{sys1}) with $\mathbf{f(x)}$ analytic in $\mathbb{R}^2$ has at most a finite number of limit cycles.
\end{theorem}
\begin{theorem}\label{poin}
{\bf(Poincar\'{e} \cite{perko}):} A planar analytic system (\ref{sys1}) cannot have an infinite number of limit cycles which accumulate on a cycle of (\ref{sys1}).
\end{theorem}
\begin{theorem}\label{poinb}
{\bf (The Generalized Poincar\'{e}-Bendixson Theorem \cite{perko}):} Suppose that $\mathbf{f(x)} \in \mathcal{C}^1(E)$ where $E$ is an open subset of $\mathbb{R}^2$. Then the system (\ref{sys1}) has only a finite number of critical points, it follows that $\omega(\Gamma)$ is either a critical point of (\ref{sys1}), a period orbit of (\ref{sys1}), or that $\omega(\Gamma)$ consists of a finite number of critical points, $\mathbf{\tilde p_1}, \mathbf{\tilde p_2}, \cdots \mathbf{\tilde p_m},$ of (\ref{sys1}) and a countable number of limit orbits of (\ref{sys1}) whose $\alpha$ and $\omega$ limit sets belong to \{$\mathbf{\tilde p_1}, \mathbf{\tilde p_2}, \cdots \mathbf{\tilde p_m}$\}.
\end{theorem}

In the following (section \ref{center}), we provide some of our newly developed theories on Moffatt vortices which will be utilized for proving our hypothesis (see section \ref{sec5}) on the finiteness of Moffatt vortices.
\subsection{Centers of Moffatt vortices: topological fixed points \& its neighborhood}\label{center}
\begin{lemma}\label{bft_vort}
The centers of the members in the sequence of Moffatt vortices are fixed points.
\end{lemma}
\noindent
{\it Proof. }
Consider the $i$-th member $V_i$ in the sequence of Moffatt vortices whose center is at ${C}_i$. When the flow reaches steady state, vortices will not deform. In other words at steady state, if we consider a fluid particle on a specific streamline then the particle will always move along that specific streamline. As such, in steady state we can define a map $\mathcal{F}_i: \overset{\circ}{\tilde{V_i}}\longrightarrow \overset{\circ}{\tilde{V_i}}$, where $\overset{\circ}{\tilde{V_i}}$ is in the interior of $\tilde{V_i}$ obtained from $V_i$ by removing only the outermost streamline  and the map $\mathcal{F}_i$ defines the rotation of the vortex $V_i$. Then each streamline will remain invariant under that function. As $\mathcal{F}_i$ is a linear map, so $\mathcal{F}_i$ is a smooth map. Therefore by Brouwer Fixed-Point theorem (Theorem \ref{bft}) with $n=2$, $\mathcal{F}_i$ must have a unique fixed point which is nothing but the center of the vortex ${C}_i$.
\hfill $\square$
\begin{definition}
\emph{
Let ${C}_i$ be the center of the $i$-th vortex $V_i$ in the sequence of Moffatt vortices. If an open neighborhood $\mathcal{O}\; (\subset \tilde{V})$ of ${C}_i$ is considered, then for any $x \in \mathcal{O} (x \neq {C}_i)$, the orbit (the path followed by fluid particles around the center ${C}_i$) $\{\Phi(x,t)\}_{t \in \mathbb{R}}$ is closed (periodic).  The largest such neighborhood $\mathcal{O}$ of ${C}_i$ is defined as the {\bf Circle Cell} of the vortex $V_i$.
}
\end{definition}
\section{Proof of finiteness of Moffatt vortices}\label{sec5}
In incompressible viscous flows, all equations governing the flows are valid only under the assumption of continuum hypothesis. According to this hypothesis, the smallest volume scale under consideration is non zero. Besides, the Kolmogorov theory \cite{wal} asserts that eddies below a certain size cannot be formed. These facts clearly lean towards the existence of a finite sequence of vortices in the corner. In the following, we prove the same with concepts developed in section \ref{sec3}  through multiple approaches.

\begin{theorem}
Suppose $\Omega$ is a closed subset of $\mathbb{R}^2$ representing an enclosed domain bounded by
solid walls (or combination of solid walls and free surfaces). Then for
a steady incompressible viscous flow, for every point $p$ on the boundary including corners, any neighborhood of $p$ contains  at most a finite number of vortices.
\end{theorem}

\begin{itemize}
\item[{\bf Proof 1:}]
We established that centers of Moffatt vortices are nothing but fixed points (referred as singular points or critical points) in Section \ref{center}. By Theorem \ref{poinb}, any non-linear dynamical system can have only finite number of singular points. If we consider a neighborhood in the corner of the solid structure then the neighborhood contains finitely many singular points. So number of vortices cannot be infinite.


\item[{\bf Proof 2:}]
We have already defined the largest neighborhood of the center of a vortex as a circle cell. By Structural Classification Theorem I (Theorem \ref{sct1}), the number of circle cells must be finite. Since only one circle cell is uniquely connected with one vortex in the flow field, the number of vortices in the corner must be finite.

\item[{\bf Proof 3:}]
In the flow domain, flow separation (reattachment) is connected with Half-saddle points (boundary-saddle points) and separation is the mechanism paving the way for the formation of a new vortex. By Lemma \ref{lm_bd}, the number of separation points is finite in the flow domain. Therefore, the number of vortices is finite in the flow domain.


\item[{\bf Proof 4:}]
Let $\mathcal{V}=(V_1, V_2, \cdots)$ be the sequence of vortices in the corner and the size of the $i^{\rm th}$ vortex be $S(V_i)$ which is defined as the distance of the center of the $i^{\rm th}$ vortex from the corner. In the sequence of vortices, any two consecutive vortices maintain a fixed ratio in size  $R=S(V_{i+1}):S(V_i)$ where $R<1$ \cite{jiten16,col,moff1,moff2}. Consider the sequence $(X_n)_{n \in \mathbb{N}}$, where $X_n=S(V_1)R^{n-1}$ represents the size of the $n$-th vortex. Since, $R<1$ therefore, $X_n \to 0$ as $n \to \infty$. From the definition of the limit of a sequence from elementary analysis \cite{rudin}, we obtain
for every $\epsilon>0$, $\exists ~ n_0 \in \mathbb{N}$ such that $|X_n|<\epsilon$ for all $n \geq n_0$. In other words, given any such positive $\epsilon$, the vortices in a ``tail of the sequence" (all the members after a fixed index, $n_0$ here) will lie within a distance of $\epsilon$ from the corner.

From Kolmogorov length scale (section \ref{kol}), if we choose $\epsilon=\eta$ then there exist an index $N_{\eta} \in \mathbb{N}$ such that $|X_n|<\eta$ for all $n \geq N_{\eta}$. Consequently, the number of vortices can never exceed $N_{\eta}$, which is a finite quantity as in order to have the number of vortices more than $N_{\eta}$ , we must have vortices violating the Kolmogorov length scale, which is impossible.
\end{itemize}
\begin{lemma}
Let us define the diameter of a vortex $V$ by
$$d=\mathrm{diam}(V):=\min_{x \in \partial\tilde{V}}\{2\Vert {C}-x \Vert_2 \colon {C} \mbox{ is the center of the vortex }V\}.$$
This $d$ is always a finite positive real number.
\end{lemma}
\noindent
{\it Proof. }
Note that $\partial\tilde{V} \neq \phi$ and ${C}$ does not belong to $\partial\tilde{V}$. Therefore the set $\{2\Vert {C}-x \Vert_2 \colon x \in \partial\tilde{V}\}$ is non-empty. Further the set is bounded below as
$\Vert {C}-x \Vert_2 >0$ for all $x \in \partial\tilde{V}$. Since $\partial\tilde{V}$ being a simple closed curve is a closed set, therefore $\exists\;x_0 \in \partial\tilde{V}$ such that
\begin{eqnarray*}
\Vert {C}-x_0 \Vert_2
&=& \inf_{x \in \partial\tilde{V}}\{\Vert {C}-x \Vert_2 \colon {C} \mbox{ is the center of the vortex }V\}\\
&=& \min_{x \in \partial\tilde{V}}\{\Vert {C}-x \Vert_2 \colon {C} \mbox{ is the center of the vortex }V\} ~ > ~0.
\end{eqnarray*}
Letting $ d =2\Vert {C}-x_0 \Vert_2 $ clearly asserts that it is a finite positive real number. This completes the proof of the lemma.
\hfill $\square$

\begin{itemize}
\item[{\bf Proof 5:}]
We define diameter, $d$ of a vortex as the diameter of the largest disk which can be inscribed inside the vortex such that the center of the disk coincides with the center of the vortex.  We  term such a disk as the diametric disk. Refer to  figure \ref{fig4} for a schematic of this situation. (Note that the the sequence of disks in this figure are actually inscribed inside the boundaries of the sequence of vortices obtained from our own simulation of the flow in a 2D triangular lid-driven cavity for a creeping flow corresponding to $Re=1$.)
\vspace{-0.8cm}
\begin{figure}[!h]
\begin{center}
\includegraphics[height=12cm,angle=-90]{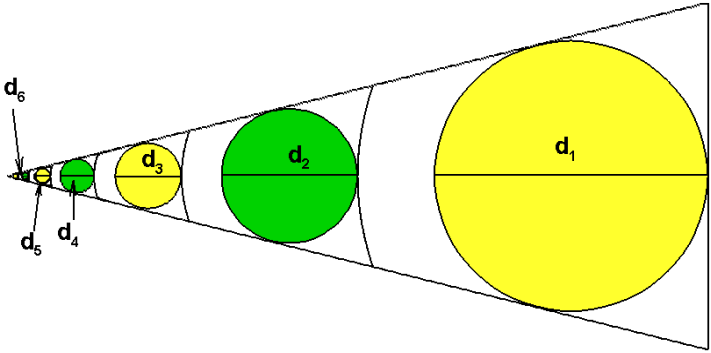}
\vspace{-5.5cm}
\caption{The diametric disks inscribed inside the vortices. The color code used here follows the alternate directions of flow inside successive vortices.}
\label{fig4}
\end{center}
\end{figure}

The above lemma clearly asserts the existence of such a length scale. Furthermore the diametric disks corresponding to a sequence of vortices are mutually disjoint as otherwise any two intersecting diametric disks will result in overlapping of two distinct vortices which is physically impossible.
Now the sequence of diameters, $(d_n)_{n \in \mathbb{N}}$ corresponding to the sequence of vortices is monotonically decreasing in nature and bounded below as they are positive quantity. So this sequence is convergent and it converges to a non-negative real number, say $\beta$.\\
{\bf Case-I:} when $\beta \neq 0$, then the total length required to accommodate all those vortices in the flow domain is greater or equal to
$ \sum_{n \in \mathbb{N}}{d_n}$. If the sequence of vortices is infinite and we replace $d_n$ by  $\beta$ (limiting diameter) in the summation, we have $ \sum_{n \in \mathbb{N}}{d_n}>\sum_{n \in \mathbb{N}}{\beta}=\infty$. This is impossible, as size of the fluid flow domain is finite. Therefore number of vortices in the flow domain cannot be infinite if $\beta \neq 0$.\\
{\bf Case-II:} When $\beta = 0$, then in order to have infinite number of vortices in the flow domain, the diameter of the extreme smallest vortex has to drop below the Kolmogorov length scale (see section \ref{kol}), $\eta \;(>0)$ , feasible length scale to measure fluid vortices. Now as $d_n \to 0$, this implies $\exists\; n_\eta \in \mathbb{N}$ such that $|d_n|<\eta$ for all $n \geq n_{\eta}$. Consequently, we can never have a vortex with diameter $d_n$ for any $n \geq n_{\eta}$. Therefore, the maximum possible number of vortices becomes less than $n_\eta$, which is finite.

Therefore, the number of vortices in the flow domain must be finite.

\item[{\bf Proof 6:}]
This proof is based on the concept of limit cycles. In order to have a clear understanding of limit cycles present in incompressible viscous flows, we exhibit certain results from our own simulation of the 3D lid-driven cavity flow.
In 3D flows, vortices are formed swirling around a three dimensional space curve known as the vortical coreline. Such a scenario can be seen from figure  \ref{fig5} where we present the vortical structure around
\begin{figure}[!h]
\begin{center}
\includegraphics[height=6cm]{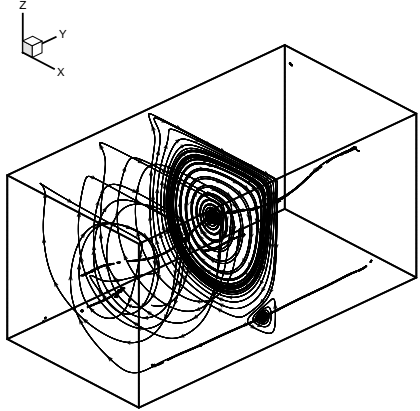}
\caption{Streamlines at the plane of symmetry for the 3D lid-driven cavity flow at $Re=1000$. One can actually see the vortical structures swirl around the vortical corelines in the figure.}\label{fig5}
\end{center}
\end{figure}
the vortical coreline from our own simulation of the flow for $Re=1000$.
The trace of limit cycles formed by the streamlines can be found in the normal plane in the context of Frenet trihendron \cite{docarmo, tyson} formed by the vectors $\hat{t}$, $\hat{n}$ and $\hat{b}$ as shown in figure \ref{fig6}, which is nothing but the plane spanned by the vectors $\hat{n}$ and $\hat{b}$. Note that $\hat{t}$ is the tangent vector to the vortical coreline at the point at which the Frenet trihendron is considered.
\begin{figure}[!h]
\begin{center}
\includegraphics[height=6cm]{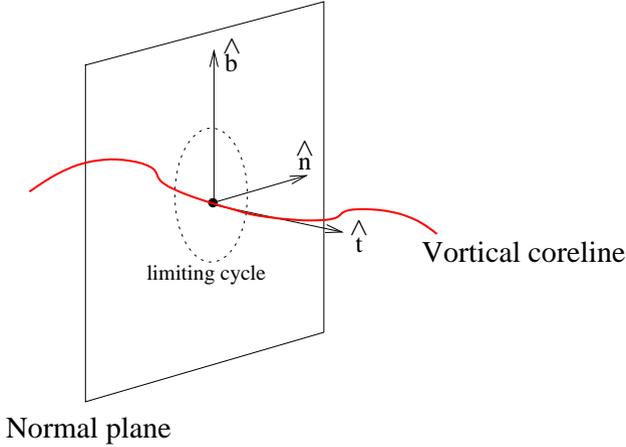}
\caption{Schematic of limit cycle in the normal plane.}\label{fig6}
\end{center}
\end{figure}
 As has been mentioned earlier, 2D flows in confined domains are in fact  idealizations of 3D flows; for example, the flow in the plane of symmetry (cross-flow plane) for the 3D lid-driven cavity.

The projection of the streamlines of the normal plane over the cross-flow plane are topologically equivalent (as projection map from plane to plane is a homeomorphism). Therefore limit cycles must be present in the cross-flow plane as well. Note that these limit cycles correspond to the vortices with centers at the coreline for planner flows in 2D.
We present two such limit cycles in the cross-flow plane corresponding to the primary vortex and the secondary vortex at the bottom right corner depicted by dark-black closed curves in figures \ref{fig7}(a) and \ref{fig7}(b) respectively.
\begin{figure}[!h]
\centering
\begin{tabular}{cc}
\hspace{-1.0cm}\epsfig{file=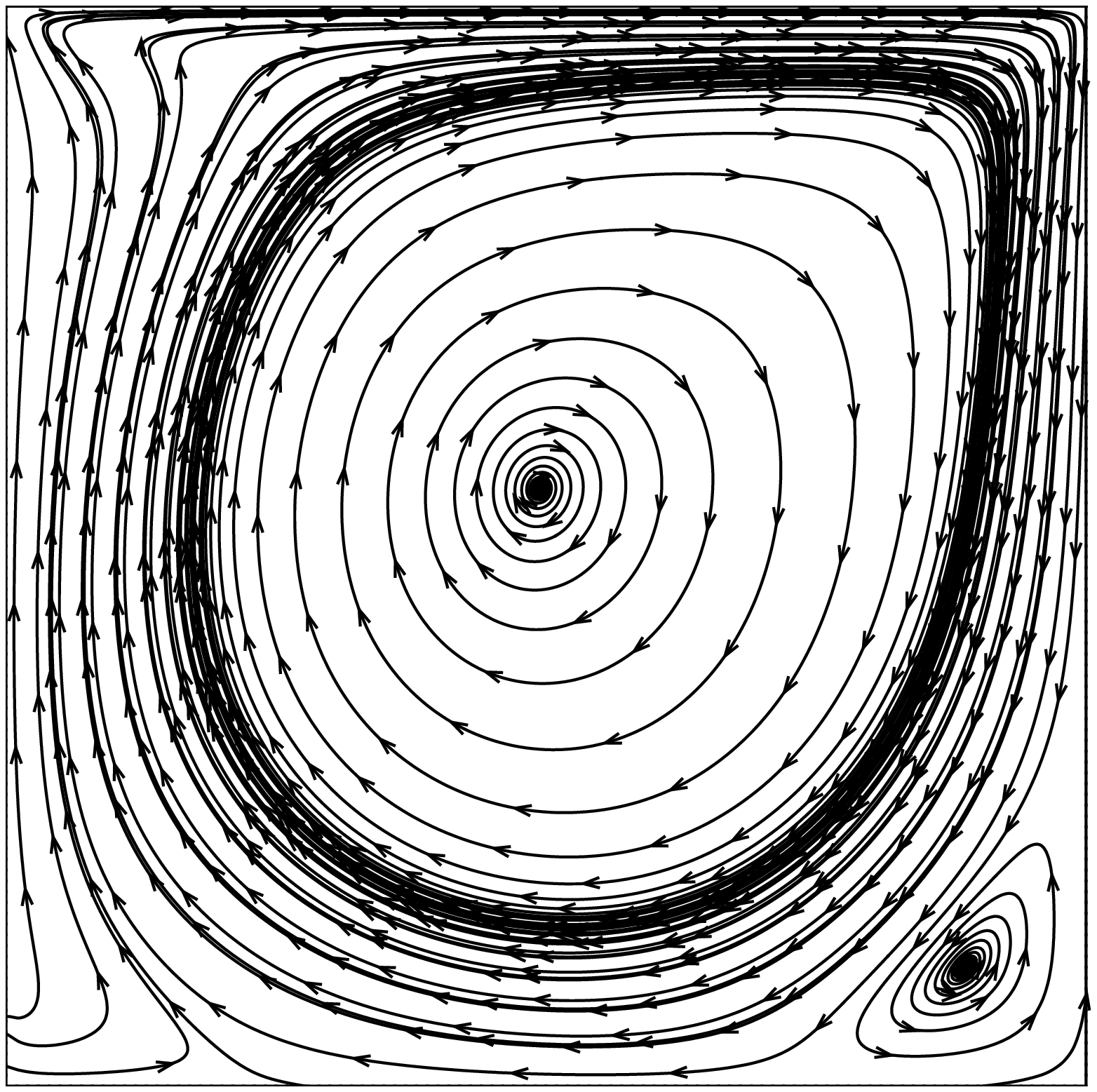,width=0.5\linewidth,clip=}
&
\hspace{-0.1cm}\epsfig{file=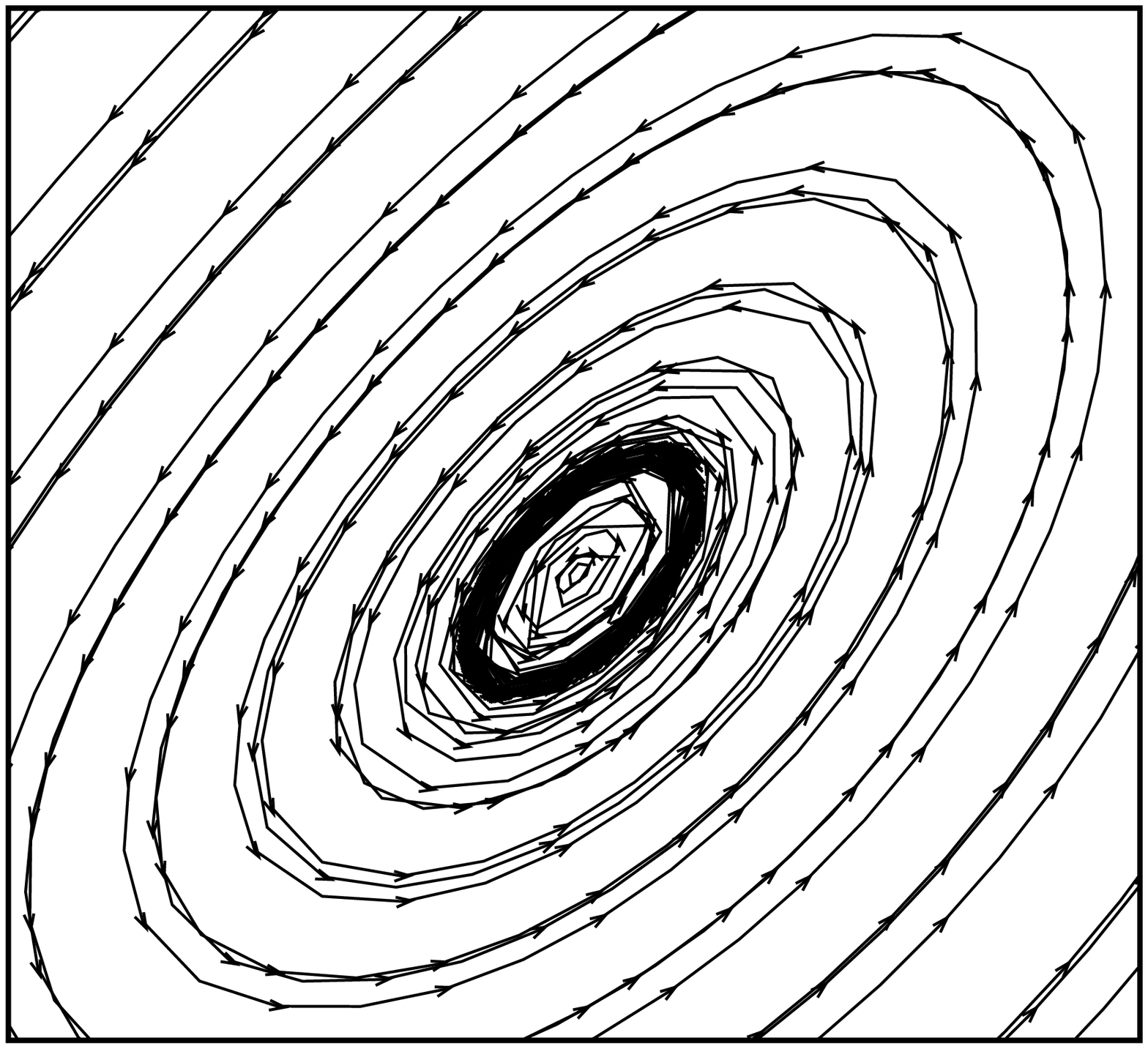,width=0.5\linewidth,clip=}
\\
(a) & (b)
\end{tabular}
\caption{Limit cycles (stable) in the lid-driven cavity flow for $Re=1000$ in the cross-flow plane: (a) Primary vortex (b) secondary corner vortex.}\label{fig7}
\end{figure}

In 3D flows, all the vortices correspond to stable limit cycles originating in foci while in 2D flows, they are simply {\it centers} (in dynamical sense) with the streamlines encircling the vortex centers. Thus each limit cycle gives rise to a vortex in the flow field. Therefore by theorems (\ref{dulac}), (\ref{poin}) and (\ref{poinb}) we conclude that number of vortices in the flow field must be finite.
\end{itemize}
\section{Conclusion}\label{sec6}
In existing literature, the occurrence of Moffatt vortices has always been synonymous with the existence of an infinite sequence. Despite the continuum hypothesis providing the base for the governing equations for incompressible viscous flows and the concept of Kolmogorov length scale in vogue, the issue of the finiteness of such sequence has continued to remain unattended. In an effort towards addressing this issue, firstly we have listed the concerns and the pertinent questions on the notion of infiniteness of such sequences and pinpointed where  the assumptions of the existing hypothesis could have gone wrong. Next, we have provided a concrete mathematical basis for predicting the correct physical phenomenon.  In order to do so, we have quantified the centers of vortices as fixed points through Brouwer fixed point theorem and further defined boundary of a vortex as a circle cell. With the aid of our newly developed theories and some existing ones, and clubbing them with elementary mathematical analysis, we proved that the number of vortices in solid corners in a bounded domain cannot be infinite. We arrived at the same conclusion by analyzing the hypothesis through six different approaches. Thus the sequence of Moffatt vortices in fluid flows around solid corners must be finite. Our observations are consistent with the recent developments of geometric theories of incompressible viscous flows.



%

\end{document}